\documentclass{ws-procs9x6}

\newcommand{\beq}{\begin{equation}}
\newcommand{\eeq}{\end{equation}}
\newcommand{\bea}{\begin{eqnarray}}
\newcommand{\eea}{\end{eqnarray}}
\newcommand{\nn}{\nonumber\\}

\newcommand{\C}{{\mathcal C}}
\newcommand{\Ord}{{\mathcal O}}
\newcommand{\R}{{\mathcal R}}
\newcommand{\Tr}{{\rm Tr}}
\newcommand{\e}{{\rm e}}
\newcommand{\bx}{{\bf x}}
\newcommand{\bp}{{\bf p}}
\newcommand{\bq}{{\bf q}}
\newcommand{\bk}{{\bf k}}
\newcommand{\br}{{\bf r}}

\begin{document}

\title{Progress in nonequilibrium\\ quantum field theory II
\footnote{\uppercase{P}repared for \uppercase{SEWM}2004. 
\uppercase{B}ased on an invited talk by 
\uppercase{J.S.} and a talk by \uppercase{J.B.}}}

\author{\vspace*{-0.2cm}J\"urgen Berges\footnote{email: 
j.berges@thphys.uni-heidelberg.de}$\,\,$
and$\,$ Julien Serreau\footnote{email: serreau@thphys.uni-heidelberg.de}}

\address{Institut f\"ur Theoretische Physik, Universit\"at Heidelberg\\ 
Philosophenweg 16, 69120 Heidelberg, Germany}
\maketitle

\abstracts{We review recent developments for the description
of nonequilibrium quantum fields, continuing hep-ph/0302210.\cite{SEWM02}}

\section{Motivations}

Nonequilibrium quantum fields play a central role in various areas 
of physics, such as cosmology, ultra-relativistic heavy-ion collisions, 
or condensed matter physics. A timely example in cosmology concerns 
the issue of reheating\cite{preheat1} of the early universe: At the end 
of inflation the universe is left in a frozen albeit very high-energy 
state, very different from the subsequent 
hot universe observed from the cosmic 
microwave background. This amounts to an enormous entropy growth due
to particle production. It is important progress that questions such 
as the far-from-equilibrium particle production 
at the end of the inflationary universe, including the subsequent process of 
thermalization, can be addressed in quantum field theory from first 
principles.\cite{Berges:2002cz} The late-time approach to quantum thermal 
equilibrium represents a complex collective phenomenon, which can occur 
on time scales dramatically larger than typical relaxation 
times.\cite{Berges:2000ur,Berges:2002wr,Aarts:2001qa} For phenomenological
applications it is therefore often crucial that different quantities
effectively thermalize on different time scales and a complete 
thermalization of all quantities may not be necessary. For instance,
an approximately time-independent 
equation of state $p=p(\epsilon)$, characterized by an almost 
fixed relation between pressure $p$ and energy density $\epsilon$, 
can form very early -- even though the system is still 
far from equilibrium. Such prethermalized quantities approximately
take on their final thermal values already at a time
when the occupation numbers of individual momentum 
modes still show strong deviations from the late-time
Bose-Einstein or Fermi-Dirac distribution.
Prethermalization is a universal 
far-from-equilibrium phenomenon which occurs on time scales 
dramatically shorter than the thermal equilibration time.\cite{Berges:2004ce} 
In order to establish such a behavior it is crucial to be able 
to compare between the time scales of prethermalization
and thermal equilibration. Approaches based on small deviations 
from equilibrium, or on a sufficient homogeneity in time underlying 
kinetic descriptions, are not applicable in this case to describe 
the required ``link'' between the early and the late-time behavior. 

Similar issues arise in the study of high-energy nuclear collisions, 
where highly excited matter is produced far from equilibrium. 
The experiments seem to indicate 
early thermalization whereas the present theoretical understanding of 
QCD suggests a much longer thermal equilibration time,\cite{Serreau:2002yr} 
thereby questioning for instance
the picture of a weakly interacting partonic gas.\cite{Shuryak:2004kh} 
This also points to the need to go beyond kinetic approaches, 
which do not provide a valid description on very short time scales. 
Efforts to go beyond kinetic descriptions include studies of the
non-equilibrium dynamics of a chiral quark-meson 
model,\cite{Berges:2002wr,Berges:2004ce} or
photon production\cite{Serreau:2003wr} in heavy 
ion collisions.

The recent progress in the theoretical description of nonequilibrium 
quantum fields is based on efficient functional 
integral techniques, so-called $n$-particle irreducible effective actions,
for which powerful nonperturbative approximation
schemes are available.\cite{Berges:2004yj} The techniques provide
an efficient approximation scheme also for situations close to, or in 
equilibrium where resummations are needed. This includes 
e.g.\ the calculation of transport coefficients\cite{Aarts:2004xs}
and thermodynamic quantities\cite{Berges:2004hn} in hot 
field theories or the study of critical phenomena near second-order 
phase transitions.\cite{Alford:2004jj} These are promising systematic
approaches to cure the convergence 
problems of perturbative calculations of thermodynamic 
quantities at high temperature such as pressure or 
entropy.\cite{Blaizot:2003tw} 

Though we focus on particle physics and
cosmology applications, we emphasize that these techniques 
can be equally applied to other nonequilibrium phenomena
in complex many body systems.
A prominent example is the study of Bose-Einstein condensates. 
The interpretation of various recent experiments, 
for instance concerning Bose condensation of molecules,\cite{CDTW} 
indicate important contributions
beyond standard mean-field type approximations
for the real-time dynamics of 
correlations.\cite{KGB} These can be systematically
taken into account using $n$PI effective action techniques.\cite{schmidt}

Apart from these applications, there has been much progress in
understanding formal aspects of approximation schemes based on 
$n$PI effective actions, such as renormalization and the validity 
of Ward identities for gauge theories. In the following we review
these developments.

\section{Non-equilibrium quantum field theory: methods}

\subsection{Standard approximations 
fail out of equilibrium}

Nonequilibrium dynamics requires the specification of an initial state 
at some time $t=0$, characterized by an initial density matrix $\rho_D$. 
All the information 
about the time-evolution of the system is encoded in the correlation 
functions of field operators $\hat\varphi$, which may be obtained from 
the generalized generating functional:
\beq
\label{functional}
 Z[R_1,R_2,\ldots;\rho_D] =
 {\rm Tr}\left\{\rho_D\,{\rm e}^{i\sum_n R_n\cdot\hat\varphi^n}\right\}\,,
\eeq
with $n$-point classical sources $R_n\equiv R_n(x_1,\ldots,x_n)$.
Here,
$
 R_n\cdot\hat\varphi^n\equiv\int_{x_1\ldots x_n}
 R_n(x_1,\ldots,x_n)\,\hat\varphi(x_1)\cdots\hat\varphi(x_n)
$,
with $\int_x \equiv \int_\C dx^0\int d^3x$. 
The closed time contour $\C$,\cite{Schwinger:1960qe} 
runs from the initial time $t=0$, where the system has been prepared, 
to an arbitrarily large time, and backward to the initial time.
The functional (\ref{functional}) can be given the following path-integral 
representation:
\beq
\label{path}
 Z[R_1,R_2,\ldots;\rho_D]=\int d\varphi_0\,d\varphi_0'\,
 \langle\varphi_0|\rho_D|\varphi_0'\rangle
 \int_{\varphi_0}^{\varphi_0'} {\mathcal D}\varphi \,
 {\rm e}^{i\left({\mathcal S}[\varphi]+\sum_n R_n\cdot\varphi^n\right)}
\eeq 
where ${\mathcal S}[\varphi]$ is the classical action and 
$\int_{\varphi_0}^{\varphi_0'} {\mathcal D}\varphi$ represents
the integral over classical paths along the contour $\C$, with fixed
boundaries on both ends at $t=0$ corresponding to the field configurations
$\varphi_0(\bx)$ and $\varphi_0'(\bx)$. One finally performs the appropriate
average over possible initial conditions. All correlation functions can be 
given similar expressions by functional differentiation. A direct evaluation 
of the corresponding path integrals in real time is currently 
not possible with 
standard Monte-Carlo techniques. Equivalently, the solution of the real-time
dynamics of the functional Schr{\"o}dinger equation is prohibitively
difficult. A frequently employed strategy is to concentrate on classical 
statistical field theory instead, which can be simulated.
This can give important insights when the number of field quanta per mode 
is sufficiently large such that quantum fluctuations 
are suppressed compared to statistical fluctuations.
However, classical Rayleigh-Jeans divergences and the lack of genuine quantum 
effects -- such as the approach to quantum thermal equilibrium
characterized by Bose-Einstein or Fermi-Dirac statistics -- limit
their use.
A coherent understanding of the time evolution in quantum field theory
is required -- a program which has made substantial progress in recent 
years with the development of powerful theoretical techniques.

There exist various powerful approximation methods 
for vacuum field theory as well as for thermal equilibrium at nonzero 
temperature. However, most standard field theoretical techniques are not 
applicable for out-of-equilibrium situations, where one encounters 
additional complications. The first new aspect concerns secularity:
The perturbative time evolution suffers from
the presence of spurious, so-called secular terms, which grow with 
time and invalidate the expansion even in the presence of a weak 
coupling. Here it is important  
to note that the very same problem appears as well for nonperturbative
approximation schemes such as standard $1/N$ expansions, where
$N$ denotes the number of field components.\footnote{Note 
that restrictions to mean-field type approximations such as
leading-order large-$N$ are insufficient. 
They suffer from the presence of an infinite number of spurious 
conserved quantities and are known to fail to describe
thermalization. Secularity enters the required next-to-leading
order corrections and beyond.} 
Typically, secularity is a not a very difficult problem and for a given
approximation there can be various ways to resolve it. 
There is a requirement, however, which poses very strong restrictions 
on the possible approximations: Universality, i.e.~the 
insensitivity of the late-time behavior to the details of the 
initial conditions. If thermal equilibrium is approached then the 
late-time result is universal in the sense that it becomes uniquely 
determined by the energy density and further conserved charges.
To implement the necessary nonlinear dynamics is demanding.
Both requirements of a non-secular and universal behavior can indeed
be fulfilled using efficient functional integral techniques:
so-called $n$-particle irreducible effective actions, where
the simplest example is given by the two-particle-irreducible 
(2PI) effective action,\cite{Cornwall:1974vz} which is described 
in the following.

\subsubsection{Secularity}
\label{sec:secular}

Here, we discuss the well-known secularity problem in the context of a 
relativistic real scalar field theory with potential 
$m^2\varphi^2/2+\lambda \varphi^4/4!$. For the argument it is
irrelevant that we assume spatial 
homogeneity and isotropy as well as $Z_2$-symmetry, so that
$\Tr\{\rho_D\,\hat\varphi(x)\}=0$, with Gaussian initial
density matrix $\rho_D$.

The time-ordered propagator along the contour $\C$ is defined as:
\beq
 G(x,y)=\Tr\{\rho_D\,T_\C(\hat\varphi(x)\,\hat\varphi(y))\}
 =\Theta_\C^+\,G^>(x,y)+\Theta_\C^-\,G^<(x,y)\,,
\eeq
where $\Theta_\C^\pm\equiv\Theta_\C(\pm(x^0-y^0))$.
One obtains to $\Ord(\lambda^2)$:
\beq
\label{twoloop}
 G_{\rm 2-loop}(x,y)=G_0(x-y)
 +\int_{uv}G_0(x-u)\,\Sigma_2(u-v)\,G_0(v-y)\,,
\eeq
where the bare inverse propagator is given by 
$G_0^{-1}(x,y)=i(\square+m^2)\delta_\C(x-y)$ and where 
$\Sigma_2$ denotes the self-energy to two-loop order: 
$\Sigma_2(x-y)=-i\lambda G_0(x,x)\delta_\C(x-y)/2-\lambda^2G_0(x-y)/6$. 
The first (one-loop) contribution corresponds to a 
simple mass shift and plays no role for the present argument.
We ignore it as well as all local insertions in the following 
discussion. 
We therefore write:
\beq
\label{lambda2}
 \Sigma_2(x-y)=-\frac{\lambda^2}{6} G_0^3(x-y)\,.
\eeq

For illustrational purposes, we introduce an effective
particle number $n_p(t)$, defined as:
\beq
\label{pnumber}
 \omega_p\left(2n_p(t)+1\right)=\left[\partial_t\partial_{t'} G^<(t,t';p)
 +\omega_p^2 G^<(t,t';p)\right]_{t'=t}\,,
\eeq
where $\omega_p=\sqrt{p^2+m^2}$ and where $G(t,t';p)$ denotes the 
spatial Fourier transform of the propagator. Writing explicitly 
the time integrals in (\ref{twoloop}), one easily
obtains, for the two-loop contribution to the particle number:
\beq
 n^{\rm 2-loop}_p(t)-n_p^0=(1+n_p^{(0)})\R^<_p(t)-n_p^{(0)}\R^>_p(t)\,,
\eeq
where, introducing the Fourier transform in frequency space 
$\bar\Sigma_2^{<,>}(\omega,p)$:
\beq
\label{R2}
 \R^{<,>}_p(t)=-\frac{1}{\omega_p}\int\frac{d\omega}{2\pi}\,\,
 \bar\Sigma_2^{<,>}(\omega,p)\,\,
 \frac{1-\cos(\omega-\omega_p)\,t}{(\omega-\omega_p)^2}\,.
\eeq
In the large-time limit, one gets:
\beq
\label{largetime}
 \R^{<,>}_p(t)\simeq \bar\sigma^{<,>}_p \times t\,,
\eeq
with the on-shell rates: 
$\bar\sigma^{<,>}_p=-\bar\Sigma_2^{<,>}(\omega_p,p)/2\omega_p$.
We recover the usual secular behavior of perturbation theory, which occurs 
whenever the rates in (\ref{largetime}) do not vanish on-shell.
For instance, in the present two-loop approximation, the production 
rate $\bar\sigma^<_p$ receives contributions from in-medium two-body 
scatterings $\bq+\bk\to\br+\bp$:
\beq
\label{rateexpr}
 \bar\sigma^<_p=\frac{\lambda^2}{4\omega_p}
 \int \tilde{dq}\tilde{dk}\tilde{dr}\,(2\pi)^4\delta^{(4)}(Q+K-R-P)\times
 n^0_q\,n^0_k\,(1+n^0_r)\,.
\eeq
where we introduced four-momenta as {\it e.g.} $Q^\mu\equiv(\omega_q,\bq)$ 
and $\tilde{dq}\equiv\frac{d^3q}{(2\pi)^32\omega_q}$.

A similar analysis reveals that higher-order contributions to Eq. 
(\ref{twoloop}) are secular as well. The standard solution to this 
problem corresponds to summing the infinite series of self-energy 
insertions:
\bea
\label{presum}
 G_2&=&G_0+G_0\cdot\Sigma_2\cdot G_0
 +G_0\cdot\Sigma_2\cdot G_0\cdot\Sigma_2\cdot G_0+\ldots\nonumber\\
 &=&G_0+G_0\cdot\Sigma_2\cdot G_2\, ,
\eea
where we have used an obvious schematic notation.
This corresponds to the usual resummation in terms of 
one-particle-irreducible (1PI) proper vertices. In the two-loop 
example discussed here, this leads to a non-secular time-evolution. 
However, the problem reappears for higher order contributions to the 
self-energy. For instance, the following $\Ord(\lambda^4)$ contribution
\beq
\label{lambda4}
 \frac{\lambda^4}{36} G_0^2(x-y)
 \times\int_{uv}G_0(x-u)\,G_0^3(u-v)\,G_0(v-y)
\eeq
contains the same time-integral as analyzed above 
and thus diverges at large times. Therefore, the systematic 
resolution of the secularity problem requires infinite resummations 
beyond the standard 1PI resummation scheme.

To further elaborate the discussion, we consider the sum of 
contributions (\ref{lambda2}) and 
(\ref{lambda4}), which can be written as: $\Sigma_4=-(\lambda^2/6)G_0^2\,
\{G_0+G_0\cdot\Sigma_2\cdot G_0\}$. Our previous analysis suggests to resum 
the infinite geometric series of self-energy ($\Sigma_2$) insertions in 
the term in brackets:
$
 -(\lambda^2/6)\,G_0^2\,
 \{G_0+G_0\cdot\Sigma_2\cdot G_0+\ldots\}
 =-(\lambda^2/6)\,G_0^2\,G_2
$.
Similarly, one can resum self-energy insertions on
the other two lines of the original two-loop diagram,
to arrive at the following simple -- and suggestive -- expression:
\beq
\label{simple}
 \Sigma_{[2]}=-\frac{\lambda^2}{6}\,G_2^3\,.
\eeq
To improve on this, we note that the simple expression (\ref{simple}), 
suggests the following iterative procedure to include 
higher orders without encountering new secular terms: 
From the propagator at the $n$-th iteration $G_{[2n]}$, compute
the resummed two-loop self-energy $\Sigma_{[2n]}=-\lambda^2G_{[2n]}/6$.
The next iteration $G_{[2(n+1)]}$ is then computed from
$G_{[2(n+1)]}^{-1}=G_0^{-1}-\Sigma_{[2n]}$. Starting from
$G_{[0]}=G_0$, this converges, after infinitely many iterations,  
to the following equation for $G\equiv G_{[\infty]}$:\footnote{This 
procedure is a direct generalization of the successive summations of 
so-called daisy and super-daisy diagrams, which correspond to the 
standard Hartree approximation. With present notations, the latter
corresponds to iterating the one-loop self-energy: 
$\Sigma_{[n]}(x,x')=-i\lambda G_{[n]}(x,x)\,\delta_\C(x-x')/2$. 
In the discussion presented here, we have discarded these (trivial) 
one-loop tadpole insertions for simplicity. }
\beq
\label{2pieom}
 G^{-1}=G_0^{-1}-\Sigma[G]
\eeq
with
\beq
\label{2piself}
 \Sigma[G]=-\frac{\lambda^2}{6}\,G^3\,.
\eeq
We arrive at a non-linear equation for the two-point function $G$.
The procedure described above guarantees that possible secular 
terms are automatically resummed at each step of the iteration. Equation 
(\ref{2piself}) is remarkably simple and suggests to reorganize the 
original perturbative expansion in terms of the full propagator. 
Before we discuss below that this is efficiently described in terms
of the 2PI effective action, we describe in the following that this
procedure also automatically respects late-time universality.  

\subsubsection{Universality}

The price to be paid for the above non-secular scheme is that
at some order one necessarily has to solve nonlinear equations 
without further approximations.
However, as we will see next, it is precisely the nonlinearity
which is required to be able to obtain universality
in the sense mentioned above.
We consider first Eq.~(\ref{presum}), which is more conveniently 
rewritten for initial-value problems as: 
$G_0^{-1}\cdot G_2=1+\Sigma_2\cdot G_2$. Using this
equation, evaluated along the contour $\C$, one obtains for
the time-derivative of the effective particle number $n_p(t)$:
\bea
\label{dernum}
 \omega_p\,\dot n_p(t)&=&{\rm Re}\int_0^t dt'\,\Big\{
 \Sigma_2^<(t,t';p)(i\partial_t-\omega_p)G_2^>(t',t;p)\nn
 &&\qquad\qquad\,
 -\Sigma_2^>(t,t';p)(i\partial_t-\omega_p)G_2^<(t',t;p)\Big\}\,.
\eea
To illustrate the content of this equation, we employ the following 
free-field like ansatz: $(i\partial_t-\omega_p)G_2^<(t',t;p)\to
-n_p(t)\,\e^{-i\omega_p(t'-t)}$,
and similarly for $G_2^>$ with $n_p(t)$ replaced by $1+n_p(t)$.
Using (\ref{R2})) one finds:
\beq
\label{pboltz}
 \dot n_p(t)=(1+n_p(t))\,\dot\R^<_p(t)-n_p(t)\,\dot\R^>_p(t)\,.
\eeq
Using the late-time behavior (\ref{largetime}), one finally gets:
\beq
\label{regular}
 n_p(t)\simeq n^{0}_p\,\e^{-\gamma_p\,t}
 +\frac{\bar\sigma^<_p}{\gamma_p}\Big(1-\e^{-\gamma_p\,t}\Big)
 \to\frac{\bar\sigma^<_p}{\gamma_p}\,,
\eeq
where $\gamma_p=\bar\sigma^>_p-\bar\sigma^<_p$. 
The late-time limit, indicated on the RHS, is completely determined by 
the initial particle number $n_p^{0}$,
see (\ref{rateexpr}). This dependence on initial conditions 
is rooted in the fact that Eq.\ (\ref{presum}) is a linear equation 
for the propagator. Late-time universality -- and the associated 
effective loss of information -- requires nonlinear equations.

To illustrate the importance of the resummation procedure discussed
in Sec.~\ref{sec:secular} and, in
particular, of the non-linear character of the resulting equations, 
we consider the time evolution of the particle number (\ref{pnumber})
for the approximation (\ref{2piself}). We emphasize that these
particle numbers are only employed for illustrational purposes. 
The quantum field theoretical
description for the propagator contains off-shell as well as memory effects
and is not limited to late-times.\cite{Berges:2000ur} Moreover, the functional 
formulation of this approach, based on the 2PI effective action 
described below provides a systematic framework for higher-order 
calculations.
Using Eq.\ (\ref{2pieom}), it is easy to check that, in the present
approximation, the time derivative of $n_p(t)$ has a similar form as 
in (\ref{dernum}) with the replacements $G_2^{<,>}\to G^{<,>}$ and 
$\Sigma_2^{<,>}\to\Sigma^{<,>}$. Employing a similar free-field like 
ansatz as above, one obtains the following Boltzmann-like equation 
at sufficiently large times:
\bea
\label{boltzmann}
 \dot n_p(t)&=&\frac{\lambda^2}{4\omega_p}\int\tilde{dq}\tilde{dk}\tilde{dr}\,
 (2\pi)^4\,\delta^{(4)}(Q+K-R-P)\,\nn
 &&\qquad\qquad\times\Big\{n_q(t)n_k(t)(1+n_r(t))(1+n_p(t))\nn
 &&\qquad\qquad\,\,\,\,-n_p(t)n_r(t)(1+n_k(t))(1+n_q(t))\Big\}\,.
\eea
The late-time equilibrium solution - corresponding to $\dot n_p(t)=0$ - 
is given by the Bose-Einstein distribution:
\beq
\label{equi}
 n^{\rm eq}_p=\frac{1}{\e^{(\omega_p-\mu)/T}-1}\,,
\eeq
and is only characterized by the temperature $T$ and chemical 
potential~$\mu$. The latter are uniquely determined 
by the total energy and particle number densities, which are conserved
in the present approximation.\footnote{While the former is an exactly 
conserved quantity, the latter is only conserved in the approximation 
(\ref{boltzmann}), which describes on-shell two-body elastic scatterings. 
In particular, it is not conserved by the original equation (\ref{2pieom}), 
which includes off-shell, particle-number-changing processes. When the 
latter are taken into account, the equilibrium distribution is given by 
(\ref{equi}) with a vanishing chemical potential.\cite{Berges:2000ur}} 
As expected, the late-time result is insensitive to the details of the 
initial condition. Note that, in contrast, resummation schemes based 
on local insertions, such as {\it e.g}.\ 
the so-called 2PPI loop expansion, fail to describe late-time 
thermalization at finite loop order.\cite{Baacke:2002ee} This demonstrates 
that it is crucial to go beyond local (mass) resummations.

\subsection{The 2PI effective action}

The previous analysis illustrates that a systematic way of respecting
both non-secularity as well as universality is to reorganize the
perturbative expansion in terms of the full propagator. This can
be done in an efficient way by performing a double Legendre transform 
of the functional (\ref{functional}) with respect to the linear and 
bilinear sources $R_1(x)$ and $R_2(x,y)$. One then obtains the so-called 
two-particle-irreducible (2PI) effective action,
a functional of the one and two-point functions $\phi(x)$ and $G(x,y)$, 
which can be conveniently parametrized as:\cite{Cornwall:1974vz} 
\beq
\label{2PIea}
 \Gamma[\phi,G] = S[\phi]+\frac{i}{2}\,\Tr \ln G^{-1}
 +\frac{i}{2}\,\Tr\, G_0^{-1}\cdot G+\Gamma_2[\phi,G]\,,
\eeq
where $\Gamma_2$ can be written as an infinite series of closed 2PI 
diagrams with lines corresponding to $G$ and vertices given by the 
shifted action ${\mathcal S}[\phi+\varphi]$.
Physical solutions corresponds to the stationarity conditions:
\beq
\label{2eq}
 \frac{\delta\Gamma[\phi,G]}{\delta \phi} = 0 \quad ,\quad
 \frac{\delta\Gamma[\phi,G]}{\delta G} = 0 \,.
\eeq
Using (\ref{2PIea}), the second of these equations is equivalent to
the equation of motion (\ref{2pieom}) for $G$, with self-energy given 
by: 
\beq
 \Sigma[\phi,G]=2i\frac{\delta\Gamma_2[\phi,G]}{\delta G}\,.
\eeq
Thus, one automatically obtains non-linear, self-consistent 
equations, as required to describe late-time thermalization.
This generalizes the resummation procedure described
previously.\footnote{For instance, the resummed two-loop self-energy
(\ref{2piself}) is obtained from the three-loop contribution
to the functional $\Gamma_2$: $i(\lambda^2/48)\int_{xy}
G^4(x,y)$.} Systematic approximations include a loop
or a $1/N$-expansion of the 2PI effective 
action beyond leading order.\cite{Aarts:2002dj}

Finally, the effective action $\Gamma[\phi]$, which encodes all 
the $n$-point functions of the theory, is obtained as:
\beq
\label{resum}
 \Gamma[\phi]=\Gamma[\phi,G[\phi]]\,,
\eeq
where $G[\phi]$ is the solution of the second of Eqs.\ (\ref{2eq}).
Equation (\ref{resum}) is an exact relation between the 1PI and 2PI 
functionals which merely states that both are equivalent representations
of the theory in the absence of higher-than-linear sources $R_{n\ge 2}$.
However, systematic ({\it e.g}.\ loop or $1/N$) expansions of the 1PI 
and 2PI functionals do not coincide order by order in general.
Equation (\ref{resum}) therefore defines a non-trivial approximation 
scheme for the computation of $n$-point functions. We call it the 
2PI resummation scheme. It is non-perturbative in nature and, as we have 
argued previously, it includes all the ingredients needed to overcome 
both the secularity and universality problems of non-equilibrium 
quantum field theory.

\subsection{Symmetries}

Linear symmetries of the classical action
are directly promoted at the level of the 2PI effective action.
For example, for a $O(N)$-symmetric scalar field theory, 
$\Gamma[\phi,G]$ is invariant under the transformation:
\beq
 \phi(x)\to\R\phi(x)\quad ; \quad G(x,y)\to\R G(x,y) \R^\dagger\,,
\eeq
where $\R$ denotes a $O(N)$ rotation. This ensures that the resummed
effective action (\ref{resum}) has the same symmetry properties as
the classical action: $\Gamma[\phi]=\Gamma[\R\phi]$. 
Notice that this holds for any 2PI approximation which respect the 
symmetry. In turn, this directly implies that Ward identities are 
automatically satisfied by the $n$-point functions calculated 
from the resummed effective action (\ref{resum}). For instance, in 
the case of spontaneously broken symmetry, the two-point function, 
evaluated at the solution $\phi=\bar\phi$ of the first of Eqs. 
(\ref{2eq}): 
\beq
\label{2pointfunc}
 \Gamma^{(2)}_{ab}(x,y)=\left.\frac{\delta^2\Gamma[\phi]}
 {\delta\phi_a(x)\delta\phi_b(y)}\right|_{\phi=\bar\phi}\,,
\eeq
where $a,b$ are $O(N)$ indices, exactly satisfies Goldstone's 
theorem\footnote{In contrast, the solution 
$G_{ab}[\bar\phi]$ of the second of Eqs. (\ref{2eq}) violates 
Goldstone's theorem in general. However, it can be 
shown that for systematic expansion schemes, such as a 2PI loop 
or $1/N$-expansion at a given order, the functions (\ref{2pointfunc}) 
and $iG_{ab}^{-1}[\bar\phi]$ only differ by higher-order terms
and, therefore, these violations are of higher order as 
well.\cite{inpreparation}}.\cite{Aarts:2002dj,vanHees:2001ik}

We emphasize that similar arguments can be applied to arbitrary 
linear and/or affine symmetry transformations including local 
ones. For instance, for QED in linear gauges, one can show that the 
2PI resummed effective action satisfies the usual Ward identities. 
In particular, one immediately concludes that the photon polarization 
tensor calculated from it is transverse in momentum space at any 
order of a 2PI loop-expansion.\cite{inpreparation}
The generalization to nonabelian gauge theories is technically more 
involved and needs to be further investigated.

\subsection{Renormalization}

A systematic renormalization procedure of the resummed propagator  
at vanishing field, $G[\phi=0]$, has been proposed in 
Refs.~\cite{vanHees:2001ik,Blaizot:2003br} for the 2PI loop-expansion 
in scalar field theories.\footnote{For a recent application  
to the 2PI $1/N$-expansion, see \cite{Cooper:2004rs}.} This is 
based on applying a BPHZ subtraction procedure to diagrams with 
resummed propagators. In particular, it has been shown that 
the subtleties arising from self-consistent resummations 
reduce in this case to a proper renormalization of the following 
Bethe-Salpeter like equation (here, $G\equiv G[\phi=0]$):
\beq
\label{BSeq}
 \bar V(x,y,z,t)=\bar\Lambda(x,y,z,t)+\frac{i}{2}\int_{uv\bar u\bar v}
 \bar\Lambda(x,y,u,v)G(u,\bar u)G(v,\bar v)
 \bar V(\bar u,\bar v,z,t)\,,
\eeq
where
\beq
 \bar\Lambda(x,y,z,t)\equiv4\frac{\delta^2\Gamma_2[\phi=0,G]}
 {\delta G(x,y)\delta G(z,t)}\Big|_{G=G[\phi=0]}\,.
\eeq

This analysis has been 
recently generalized in \cite{Berges:2004hn,inpreparation} to include the 
field-dependence of the 2PI resummed effective action (\ref{resum}). 
This allows one to determine all the counterterms of the theory
at a given level of approximation from suitable renormalization conditions 
and to obtain finite results for arbitrary $n$-point functions.
As an explicit example, we consider the one-component scalar field theory 
discussed in the previous sections at order $\lambda_R^2$ in the 2PI 
coupling-expansion, where $\lambda_R$ denotes the renormalized coupling. 
We employ renormalization conditions for the two-point function, 
$\Gamma^{(2)}$, and four-point function, $\Gamma^{(4)}$, given by: 
\bea
\label{2point}
 \Gamma^{(2)}(x,y) &\equiv& 
 \frac{\delta^2 \Gamma[\phi]}{\delta\phi(x)\delta\phi(y)}
 \Big|_{\phi=0}\,,\\
\label{4point}
 \Gamma^{(4)}(x,y,z,t) &\equiv& 
 \frac{\delta^4\Gamma[\phi]}{\delta\phi(x) 
 \delta\phi(y)\delta\phi(z)\delta\phi(t)}\Big|_{\phi=0}\,.
\eea
Without loss of generality we use renormalization conditions
for $\phi = 0$ which in Fourier space read: 
\bea
\label{eq:r1}
 Z\,\Gamma^{(2)}(p^2)|_{p=0}&=&-m_R^2\,,\\
\label{eq:r2}
 Z\,\frac{d}{d p^2}\Gamma^{(2)}(p^2)|_{p=0}&=&-1\,,\\
\label{eq:r3}
 Z^2\,\Gamma^{(4)}(p_1,p_2,p_3)|_{p_1=p_2=p_3=0}&=&-\lambda_R \, ,
\eea
with the wave function renormalization $Z$. 
Defining the renormalized fields as:
$\phi_R = Z^{-1/2} \phi$ and $G_R[\phi_R]=Z^{-1}G[\phi]$,
and introducing the usual counterterms:
$Zm^2=m_R^2+\delta m^2$, $Z^2\lambda=\lambda_R+\delta\lambda$ and
$\delta Z=Z-1$, the classical contribution to the effective action 
(\ref{resum}) reads:
\beq
\label{classical}
 S[\phi]=\int_x\left(
 \frac{1+\delta Z}{2}\partial_{\mu}\phi_R\partial^{\mu}\phi_R
 -\frac{m_R^2+\delta m^2}{2}\phi_R^2
 -\frac{\lambda_R+\delta\lambda}{24}\phi_R^4\right)\,.
\eeq
Similarly, one can write for the one-loop part: 
$\Tr\ln G^{-1}[\phi] = \Tr\ln G_R^{-1}[\phi_R]$ 
up to an irrelevant constant, and:
\bea
\label{oneloop}
 \frac{i}{2}\Tr\, G_{0}^{-1}[\phi] G[\phi] & =& 
 -\frac{1}{2} \int_x \left[(1+\delta Z_1)\square_x
 +m_R^2+\delta m_1^2\right]G_R(x,y;\phi_R)|_{x=y}\nn
 &&-\frac{\lambda_R+\delta\lambda_1}{4}\int_x 
 \phi_R^2(x) G_R(x,x;\phi_R)\, . 
\eea
Here, $\delta Z_1$, $\delta m_1^2$ and $\delta \lambda_1$
denote the same counterterms as above,
however approximated to the given order. Finally, the contribution
from the 2PI functional $\Gamma_2$ in Eq.\ (\ref{2PIea})
reads, at order $\lambda_R^2$:
\bea
\label{higherloop}
 \Gamma_2[\phi,G[\phi]] &=&
 -\frac{\lambda_R+\delta\lambda_2}{8}\int_x G_R^2(x,x;\phi_R)
 +i\frac{\lambda_R^2}{48}\int_{x y}G_R^4(x,y;\phi_R)
 \nn
 &&+i\frac{\lambda_R^2}{12}\int_{x y}\phi_R(x)G_R^3(x,y;\phi_R)\phi_R(y)\,.
\eea

One first has to calculate the solution $G_R[\phi_R]$ of the 
stationarity condition for the 2PI effective action. For this, 
one has to impose the same renormalization conditions as for the 
propagator (\ref{eq:r1})-(\ref{eq:r2}) in Fourier space:
\bea 
\label{eq:rencond1}
 iG_R^{-1}(p^2;\phi_R)|_{p=0,\phi_R=0}&=&-m_R^2\,,\\
\label{eq:rencond2}
 \frac{d}{dp^2}iG_R^{-1}(p^2;\phi_R)|_{p=0,\phi_R=0}&=&-1\,,
\eea
together with the condition:
\beq
\label{eq:rencond3}
 \bar V_R(p_1,p_2,p_3)|_{p_1 = p_2 = p_3 = 0} = - \lambda_R \, ,
\eeq
for the renormalized ``four-point'' field, obtained from Eq. (\ref{BSeq}). 
It is easy to check that conditions (\ref{eq:rencond1})-(\ref{eq:rencond3}) 
fix all the counterterms of the effective action at vanishing field,
namely $\delta Z_1$, $\delta m_1^2$ and $\delta\lambda_2$. This 
corresponds to the case considered in \cite{vanHees:2001ik,Blaizot:2003br}. 

The remaining counterterms are to be determined from the 
renormalization conditions (\ref{eq:r1})-(\ref{eq:r3}). A similar analysis 
to the one performed in \cite{Blaizot:2003br} shows that the coupling 
divergences appearing in the two-point function (\ref{2point}) can be 
absorbed in a proper renormalization of the following four-point 
field:\cite{inpreparation}
\beq
 \label{eq:4field}
 V_R(x,y,z,t) \equiv 
 \frac{\delta^2 i G_R^{-1}(x,y;\phi_R)}{\delta \phi_R(z) 
 \phi_R(t)}\Big|_{\phi_R=0} \,,
\eeq
which satisfies a Bethe-Salpeter like equation similar to (\ref{BSeq}).
The corresponding condition reads, in momentum space:
\beq
\label{eq:rencond4}
 V_R(p_1,p_2,p_3)|_{p_1 = p_2 = p_3 = 0} = - \lambda_R \, .
\eeq
The latter, together with Eqs.\ (\ref{eq:r1})-(\ref{eq:r2}) fix the 
counterterms associated with the quadratic field-dependence in Eqs. 
(\ref{classical})-(\ref{higherloop}), namely
$\delta Z$, $\delta m^2$ and $\delta\lambda_1$.\footnote{In the present
approximation, one has $\delta Z_1=\delta Z$ and $\delta m_1^2=\delta m^2$, 
which follow from the identity $\Gamma^{(2)}(x,y)=iG^{-1}(x,y;\phi=0)$, 
as well as $\delta\lambda_2=\delta\lambda_1$, 
which follows from $V(x,y;z,t)=\bar V(x,y,z,t)$.\cite{Berges:2004hn} 
Notice that, although 
these identities are true in the exact theory, they may not be satisfied 
in general at the level of approximations.} 
Finally, the remaining counterterm $\delta\lambda$, associated with
the quartic field-dependence, is determined from Eq.\ (\ref{eq:r3}).

It is important to emphasize that Eqs.\ (\ref{eq:rencond1}) 
and (\ref{eq:rencond2}), as well as (\ref{eq:rencond3}) and 
(\ref{eq:rencond4}) are not independent conditions.
Instead, they are fixed by the renormalization conditions 
(\ref{eq:r1})-(\ref{eq:r3}). This express the fact that in the 
exact theory, the following relations between renormalized 
quantities hold:
$\Gamma_R^{(2)}=iG_R^{-1}(\phi_R=0)$ and
$\Gamma_R^{(4)}=\bar V_R=V_R$. Thus, the renormalization 
procedure described here provides an efficient fixing of all the above
counterterms, including those associated with the explicit 
field-dependence of the effective action. The latter play a crucial 
role {\it e.g}.\ in the broken phase and are essential for the 
determination of the effective potential.

\subsection{Equivalence hierarchy for $n$PI effective actions}

The 2PI resummation scheme provides a self-consistent dressing of
the field $\phi$ and the propagator $G$. This can be generalized to 
higher $n$-point functions as well: The $n$PI effective 
action $\Gamma[\phi,G,V_3,\ldots,V_n]$ provides a self-consistent 
description for the dressed proper vertices $V_3,\ldots,V_n$, 
which are obtained the stationarity conditions: 
$\delta\Gamma/\delta V_3=0,\ldots,\delta\Gamma/\delta V_n =0$.
As before, non-secular approximation respecting late-time 
universality can be obtained from systematic expansion schemes 
of higher effective actions.
In general this leads to non-linear integro-differential equations
for the $n$-point functions, which are now independent self-adjusting 
variables. As described previously, this is a crucial ingredient for 
late-time universality. 

The use of $n$PI effective actions with $n > 2$ can be important to 
describe non-Gaussian initial density 
matrix for non-equilibrium systems. They are also relevant in the 
context of high-temperature gauge theories for a quantitative description 
of transport coefficients beyond the ``leading-log'' 
approximation,\cite{Berges:2004pu} or in the context of critical 
phenomena near second-order phase transitions. For instance, the 
quantitative description of the critical exponents of scalar $\phi^4$ 
theory goes beyond any finite order 2PI 
loop-expansion\footnote{Critical phenomena 
can be described 
using the 2PI $1/N$-expansion beyond LO.\cite{Alford:2004jj}}.\cite{Alford:2004jj} It requires taking into account vertex 
corrections that start with the 4PI effective action to four-loop
order.

For practical purposes, a crucial observation is 
that there exists an equivalence hierarchy between $n$PI
effective actions. For instance, in the case of a
loop expansion one has:\cite{Berges:2004pu}
\bea
 \Gamma^{\rm (1loop)}[\phi]&\ne&\Gamma^{\rm (1loop)}[\phi,G]
 =\Gamma^{\rm (1loop)}[\phi,G,V_3]=\ldots\,,\nn
 \Gamma^{\rm (2loop)}[\phi]&\ne&\Gamma^{\rm (2loop)}[\phi,G]
 =\Gamma^{\rm (2loop)}[\phi,G,V_3]=\Gamma^{\rm (2loop)}[\phi,G,V_3,V_4]
 =\ldots\nn
 \Gamma^{\rm (3loop)}[\phi]&\ne&\Gamma^{\rm (3loop)}[\phi,G]
 \ne\Gamma^{\rm (3loop)}[\phi,G,V_3]=\Gamma^{\rm (3loop)}[\phi,G,V_3,V_4]
 =\ldots\nn
 \vdots\nonumber
\eea   
where $\Gamma^{{\rm (}n{\rm loop)}}$ denotes the approximation of the 
respective effective action to $n$-loop order in the
absence of external sources. 
E.g.~for a two-loop approximation
all $n$PI descriptions with $n \ge 2$ are equivalent and
the 2PI effective action captures already the 
complete answer for the self-consistent description
up to this order. In contrast, a self-consistently complete 
result to three-loop order requires at least the 3PI effective action 
in general, etc.\footnote{There exist further simplifications 
which can decrease the hierarchy even further 
such that lower $n$PI effective actions can be sufficient in
practice.  For instance, for a vanishing field $\phi$, one has,
in the absence of sources: $\Gamma^{\rm (3loop)}[\phi=0,G] = 
\Gamma^{\rm (3loop)}[\phi=0,G,V_3=0,V_4] = \ldots$.}
Typically the 2PI, 3PI or maybe the 4PI effective action  
captures already the complete answer for the self-consistent 
description to the desired/computationally feasible order of 
approximation.  

We thank the organizers for this stimulating meeting. We also thank 
Sz.~Bors\'anyi, U.~Reinosa and C. Wetterich for fruitful collaborations.

\end{document}